\pdfoutput=1

\documentclass{svproc}
\usepackage{graphicx, array, subcaption, amsfonts, float, xcolor}

\usepackage{hyperref}
\hypersetup{bookmarksnumbered=true, bookmarksopen=true}

\usepackage[absolute,showboxes,overlay]{textpos}

\begin{document}
\mainmatter              
\title{Segmentation of Blood Vessels, Optic Disc Localization, Detection of Exudates and Diabetic Retinopathy Diagnosis from Digital Fundus Images}

\titlerunning{Segmentation of Blood Vessels, ...}  
%

\author{Soham Basu\inst{1 [0000-0001-9451-3701]}, Sayantan Mukherjee\inst{2 [0000-0001-9385-7369]}, Ankit Bhattacharya\inst{2 [0000-0003-1434-7892]}, 
Anindya Sen\inst{1 [0000-0002-3770-7734]}}
\authorrunning{Soham Basu et al.}
%
\tocauthor{Soham Basu, Sayantan Mukherjee, Ankit Bhattacharya,  Anindya Sen}
\institute{Department of Electronics and Communication Engineering, Heritage Institute of Technology, Kolkata, West Bengal, India,\\
\email{soham.basu07@gmail.com, anindya.sen@heritageit.edu.in},
\and
Tata Consultancy Services, Kolkata, West Bengal, India\\
\email{mukherjee.sayantan96@gmail.com, ankub10@gmail.com},
}

\maketitle              

\begin{abstract}
Diabetic Retinopathy (DR) is a complication of long-standing, unchecked diabetes and one of the leading causes of blindness in the world. This paper focuses on improved and robust methods to extract some of the features of DR, viz. Blood Vessels and Exudates. Blood vessels are segmented using multiple morphological and thresholding operations. For the segmentation of exudates, k-means clustering and contour detection on the original images are used. Extensive noise reduction is performed to remove false positives from the vessel segmentation algorithm’s results. The localization of Optic Disc using k-means clustering and template matching is also performed. Lastly, this paper presents a Deep Convolutional Neural Network (DCNN) model with 14 Convolutional Layers and 2 Fully Connected Layers, for the automatic, binary diagnosis of DR. The vessel segmentation, optic disc localization and DCNN achieve accuracies of 95.93\%, 98.77\% and 75.73\% respectively. The source code and pre-trained model are available \textcolor{blue}{\href{https://github.com/Sohambasu07/DR_2021}{\textit{here}}}.

\end{abstract}

\section{Introduction}
\subsection{Diabetic Retinopathy}
\begin{figure}
    \centering
    \includegraphics[width=0.5\textwidth]{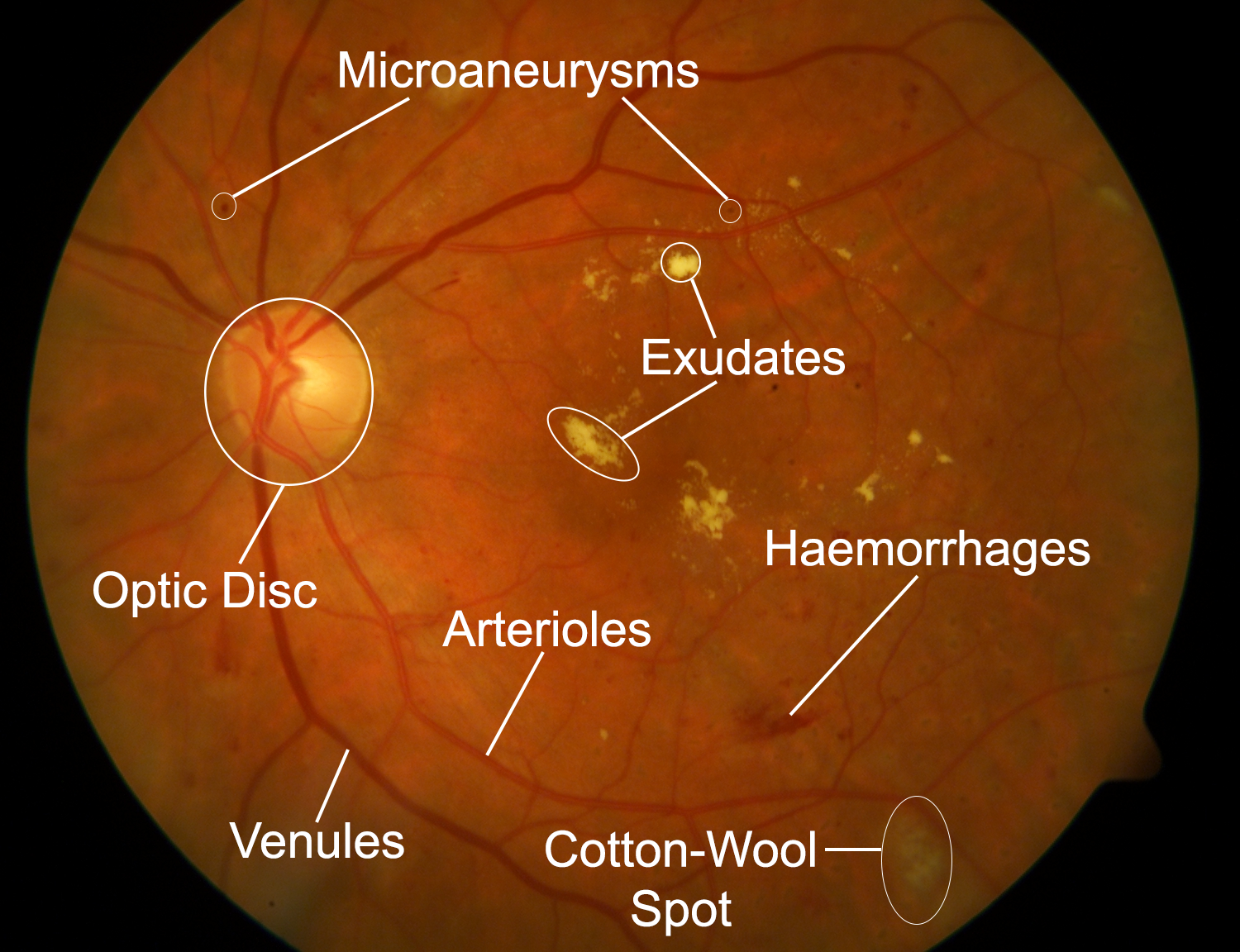}
    \caption{Different features in a typical DR affected image}
    \label{fig: fig1}
\end{figure}

Diabetic Retinopathy is a direct consequence of prolonged, unchecked diabetes, wherein the retinal blood vessels get damaged and leak fluid into the retina. If left untreated, DR can eventually lead to total blindness. DR can be classified as: Mild, Moderate, Severe and Proliferative Diabetic Retinopathy (PDR). These stages can be identified by the presence and extent of certain features (Fig. \ref{fig: fig1}).


\setlength{\TPHorizModule}{\paperwidth}\setlength{\TPVertModule}{\paperheight}
\TPMargin{5pt}

\newcommand{\copyrightstatement}{
    \begin{textblock}{0.57}(0.22,0.86)
         \noindent
         \scriptsize This is an Author Accepted Manuscript of the following chapter: Soham Basu, Sayantan Mukherjee, Ankit Bhattacharya,  Anindya Sen, Segmentation of Blood Vessels, Optic Disc Localization, Detection of Exudates and Diabetic Retinopathy Diagnosis from Digital Fundus Images, published in Proceedings of Research and Applications in Artificial Intelligence, edited by Indrajit Pan, Anirban Mukherjee, Vincenzo Piuri, 2021, Springer reproduced with permission of Springer Nature Singapore Pte Ltd. 2021. 
         The final authenticated version is available online at: \href{https://dx.doi.org/10.1007/978-981-16-1543-6\_16}{https: //dx.doi.org/10.1007/978-981-16-1543-6\_16}
         Users may only view, print, copy, download and text- and data-mine the content, for the purposes of academic research. 
         The content may not be (re-)published verbatim in whole or in part or used for commercial purposes. Users must ensure that the author’s moral rights as well as any third parties’ rights to the content or parts of the content are not compromised.
         
    \end{textblock}
}

\copyrightstatement


\subsection{Motivation}

Ophthalmologists identify Diabetic Retinopathy based on certain features viz. blood vessel area, soft and hard exudates, hemorrhages, cotton wool spots and microaneurysms. Automatic extraction of these features from fundus images help in the quick and early diagnosis of DR. Proliferative Diabetic Retinopathy is easily identified by studying the abnormal pattern of retinal blood vessels.

\subsection{Proposed Method}

The proposed algorithm utilizes the structure and contrast of the darker blood vessels with respect to the brighter background, and aims to efficiently and accurately segment the vessels from retinal fundus images.
Next, the structural profile of the Optic Disc is used to generate a template and the images are matched with this template to calculate the similarity between the two.
The exudates detection method performs k-means clustering to cluster the different intensities in the original image, and extract the pixels with the highest intensities. 
Finally, the proposed DCNN employs 14 convolutional layers to generate feature maps from images and predict the correct labels for the diagnosis of DR.

\section{Background}

Wang et al.\cite{wang} demonstrated the use of two classifiers – Convolutional Neural Network (CNN) and Random Forest (RF), which can automatically learn features from raw images and predict patterns, by combining feature learning and traditional learning. Zhang et al. \cite{zhang} proposed an algorithm which classifies vessel pixels using a texton dictionary. It focused more on the thin vessel regions which increased its sensitivity. However, non-vessel pixels may be recognized as vessel pixels, thereby decreasing accuracy and specificity. Singh and Srivastava \cite{singh} used entropy-based optimal thresholding and length filtering, while Al-Diri et al. \cite{aldiri} used active contours.

Abbadi et al. \cite{abbadi} used the grey levels of the Optic Disc (OD) to approximate its boundary. Abdullah and Fraz \cite{abdullah} used grow-cut, Mary et al. \cite{mary} used active contours and Marin et al. \cite{marin} used thresholding on morphologically transformed images. Some use the green channel, the red channel or a combination of both. These algorithms fail due to the poor contrast of the OD or saturation due to overexposure in the red channel. Besides, the shapes and sizes of exudates may often be comparable to that of the OD.

Liu et al. \cite{liu} used thresholding and region growing to detect exudates. Long et al. \cite{long} used Dynamic Thresholding and SVM Classification, while Ege et al. \cite{ege} used Bayesian, Mahalanobis and nearest neighbor classifiers for the same.

Lam et al. \cite{lam} proposed the concept of transfer learning using pre-trained neural networks like GoogLeNet and AlexNet from ImageNet. Pratt et al. \cite{pratt} proposed another CNN model which was trained using Kaggle’s DR database. However, it could only be trained on a high-end GPU to achieve acceptable results.

\section{Materials and Methods}

\subsection{Hardware and Libraries}

All the algorithms proposed in this paper were written in Python (version 3.6.9) using OpenCV and Scikit-learn libraries in Jupyter notebooks \cite{opencv}, \cite{scikit}. TensorFlow Keras was used to build the DCNN for DR diagnosis. The Jupyter notebooks were executed on virtual machines provided by Google Colaboratory (standard runtime) which consist of Intel® Xeon single core, 2.3 GHz CPUs and around 12.5 GB of available RAM. The DCNN was trained on the GPU runtime, with a single Tesla K80 GPU.

\subsection{Datasets}

For the Blood Vessel Segmentation, we have used the DRIVE dataset \cite{drive} . It is an openly available dataset consisting of 40 images of size 565 × 584 pixels, split equally among training and test sets. Thirty three out of forty images are without signs of DR and seven images are with the signs of DR. The proposed method has been evaluated on the test set and reported in this paper. The IDRiD \cite{idrid} Segmentation dataset was used for the Localization of the Optic Disc and detection of hard exudates. It is a publicly available dataset consisting of 81 images of 4288 × 2848 pixels each, split into training and tests sets of 54 and 27 images respectively. The IDRiD Disease Grading dataset was used for training and testing the DCNN for the diagnosis of Diabetic Retinopathy. It is also a publicly available dataset consisting of 516 images, split into training and test sets of 413 and 103 images respectively.

\subsection{Proposed Methods}

\subsubsection{Blood Vessel Segmentation}

\begin{figure}
     \centering
     \begin{subfigure}[b]{0.24\textwidth}
         \centering
         \includegraphics[width=0.72\textwidth]{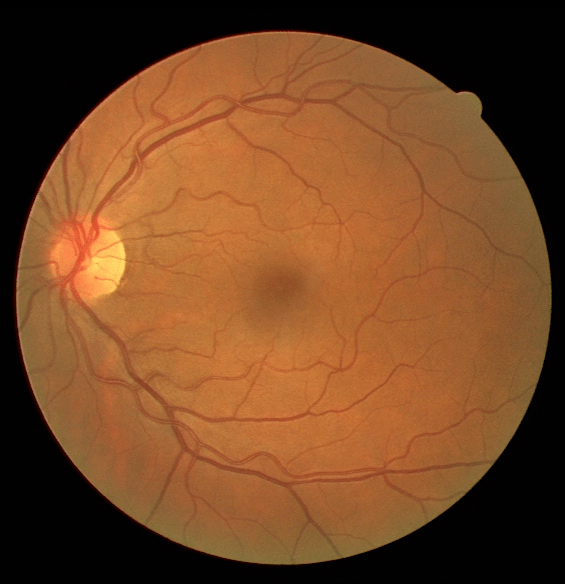}
         \caption{}
         \label{fig: fig2a}
     \end{subfigure}
     \begin{subfigure}[b]{0.24\textwidth}
         \centering
         \includegraphics[width=0.72\textwidth]{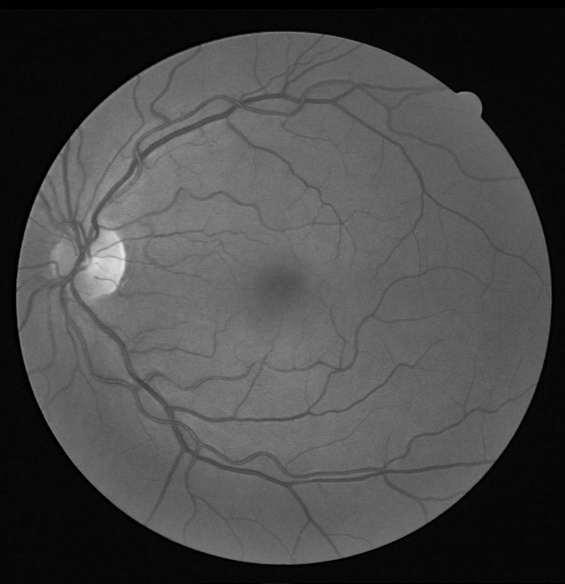}
         \caption{}
         \label{fig: fig2b}
     \end{subfigure}
     \begin{subfigure}[b]{0.24\textwidth}
         \centering
         \includegraphics[width=0.72\textwidth]{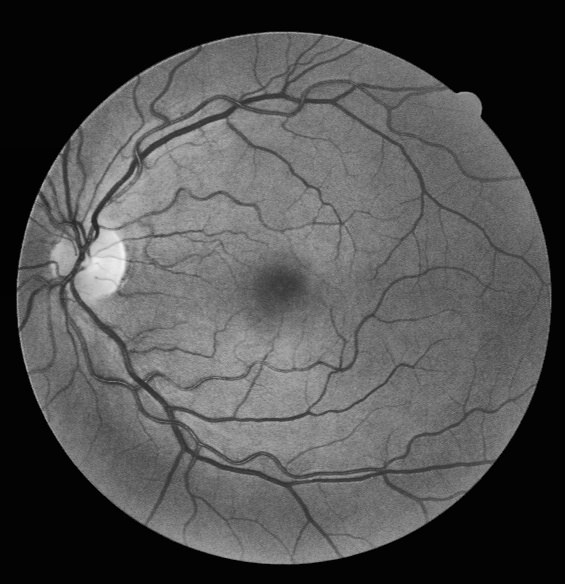}
         \caption{}
         \label{fig: fig2c}
     \end{subfigure}
     \begin{subfigure}[b]{0.24\textwidth}
         \centering
         \includegraphics[width=0.72\textwidth]{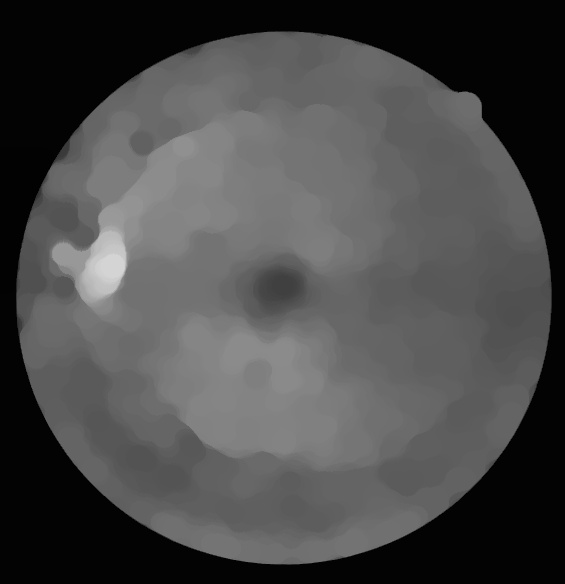}
         \caption{}
         \label{fig: fig2d}
     \end{subfigure}
     \begin{subfigure}[b]{0.24\textwidth}
         \centering
         \includegraphics[width=0.72\textwidth]{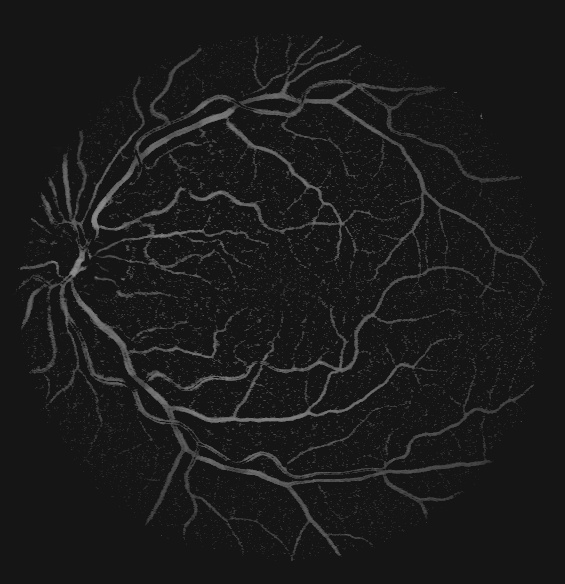}
         \caption{}
         \label{fig: fig2e}
     \end{subfigure}
     \begin{subfigure}[b]{0.24\textwidth}
         \centering
         \includegraphics[width=0.72\textwidth]{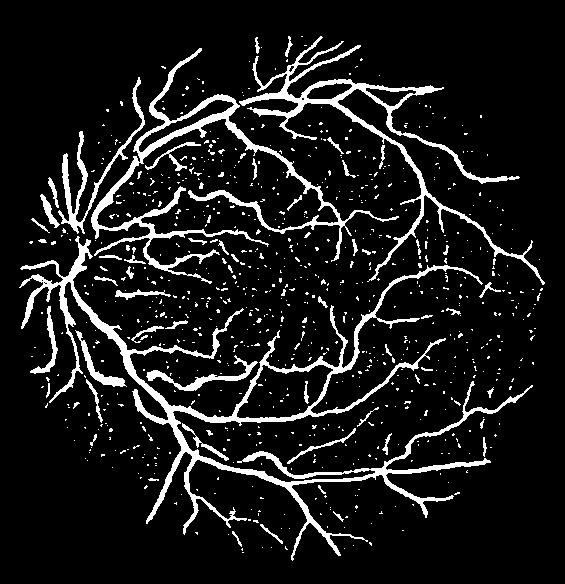}
         \caption{}
         \label{fig: fig2f}
     \end{subfigure}
     \begin{subfigure}[b]{0.24\textwidth}
         \centering
         \includegraphics[width=0.72\textwidth]{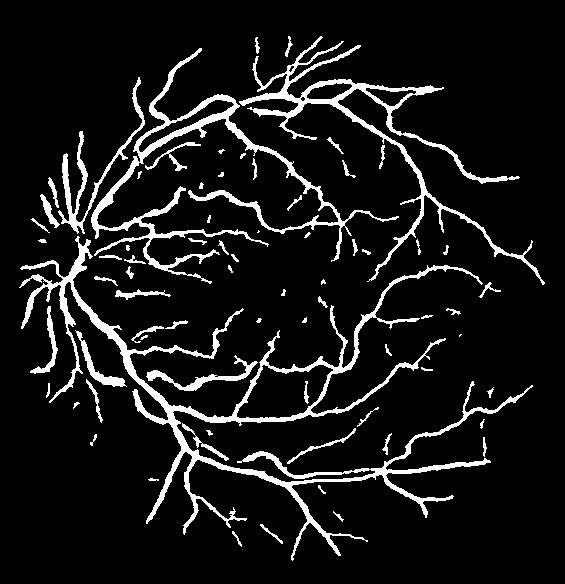}
         \caption{}
         \label{fig: fig2g}
     \end{subfigure}
     \vspace{-1mm}
     \caption{(a) Original Image, (b) Green channel component of (a), (c) CLAHE applied image, (d) Background estimated after Alternate Sequential Filtering, (e) Image (d) subtracted from (c) and CLAHE applied again, (f) Median blur and thresholding, (g) Final segmentation output.}
     \label{fig: fig2}
\end{figure}

\begin{figure}
    \centering
    \includegraphics[width=0.9\textwidth]{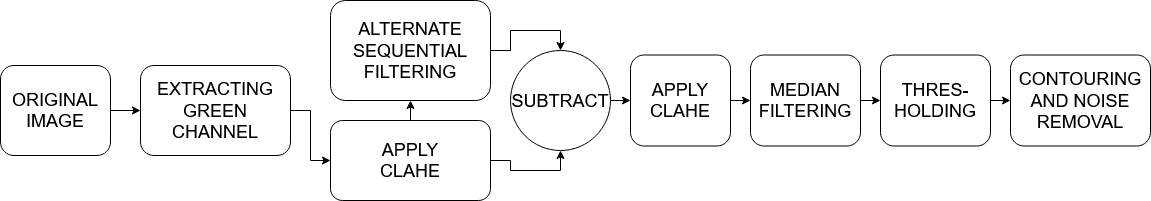}
    \caption{Blood Vessel Segmentation Flow}
    \label{fig: fig3}
\end{figure}

The green channel of the original RGB image (Fig. \ref{fig: fig2b}) was selected because the retina is most sensitive to the green wavelength of light and the vessels have the best contrast in the green channel. Contrast Limited Adaptive Histogram Equalization (CLAHE) was applied to the green channel image to create better local contrast of the vessel pixels (Fig. \ref{fig: fig2c}). The image was passed through four iterations of Alternate Sequential Filtering, where elliptical kernels of sizes (5, 5), (7, 7), (15, 15), (11, 11) were applied to the respective iterations (Fig. \ref{fig: fig2d}). The result was subtracted from the image in Fig. \ref{fig: fig2c} to generate a rough outline of the vessels, removing the background features. CLAHE was applied a second time to create an even better contrast against most other background features (Fig. \ref{fig: fig2e}). The image was then passed through a Median Filter with a kernel size of (3, 3) to filter out salt and pepper noise. Thresholding was done based on the average intensity value (Fig. \ref{fig: fig2f}). Contour detection was then performed to remove larger, isolated blobs. This yields the final segmented vessels in Fig. \ref{fig: fig2g}. The algorithm's steps are shown in Fig. \ref{fig: fig3}.

\vspace{-3mm}

\vspace{-10mm}
\paragraph{Contour Detection}

Contour detection employs Green’s theorem. If $C$ is a simple, closed curve, $D$ is the plane enclosed by $C$, $P$ and $Q$ are functions of $(x, y)$ defined on an open region containing $D$ (given that their partial derivatives exist in that region), then Green’s theorem can be stated as:

\begin{equation}
 \int_C P\,dx + Q\, dy = \int\!\!\!\int_D \left({\partial Q\over \partial x} - {\partial P\over \partial y}\right)
\end{equation}

where the path of integration along the curve $C$ is counterclockwise.

\subsubsection{Localization of Optic Disc}

\begin{figure}
     \centering
     \begin{subfigure}[b]{0.24\textwidth}
         \centering
         \includegraphics[width=0.8\textwidth]{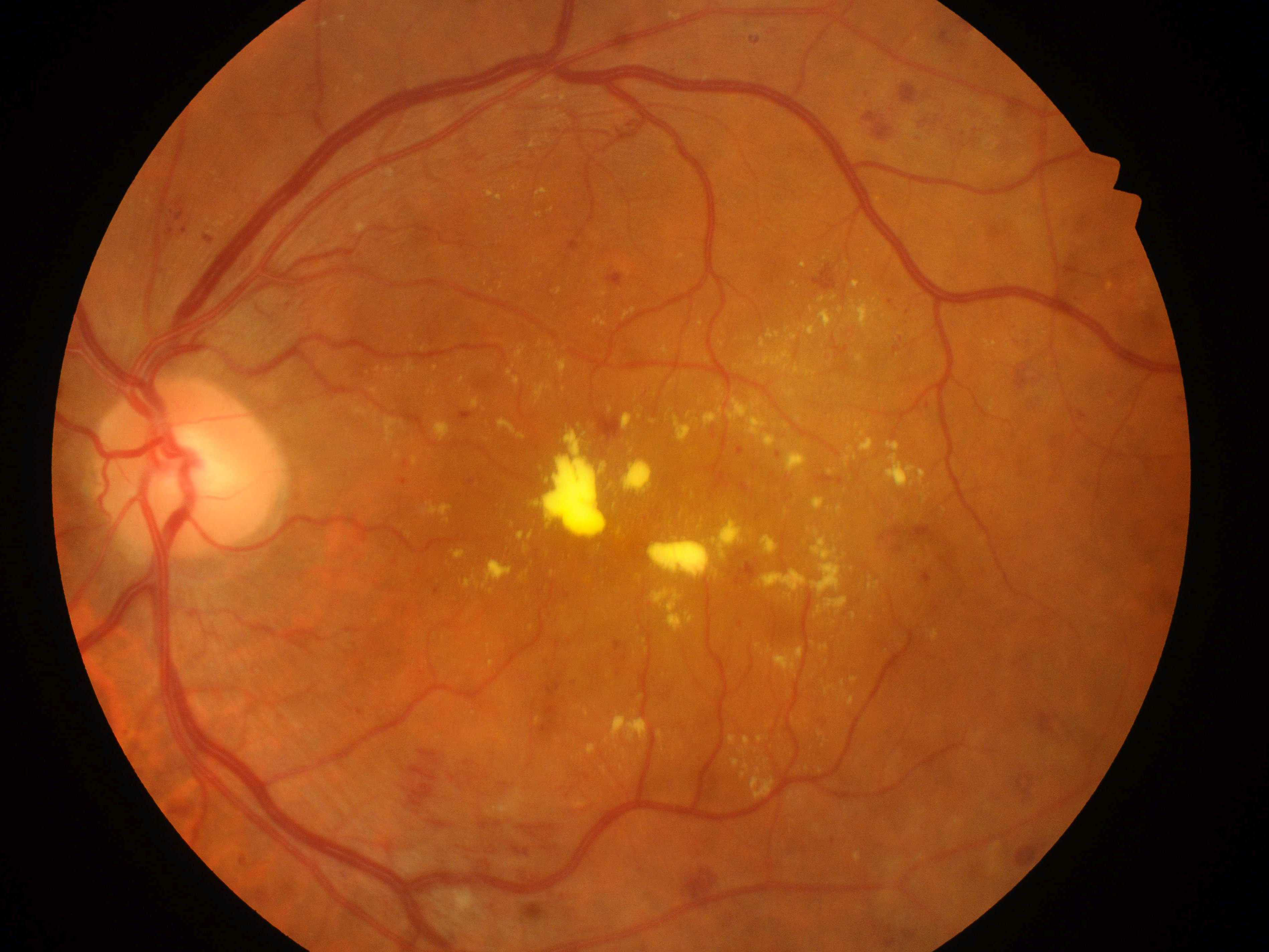}
         \caption{}
         \label{fig: fig4a}
     \end{subfigure}
     \hfill
     \begin{subfigure}[b]{0.24\textwidth}
         \centering
         \includegraphics[width=0.8\textwidth]{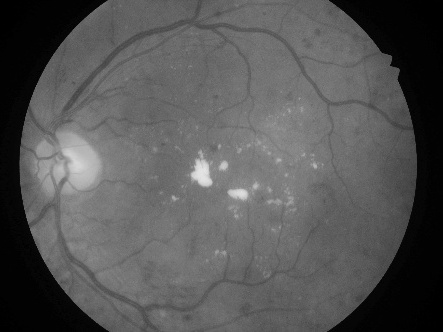}
         \caption{}
         \label{fig: fig4b}
     \end{subfigure}
     \hfill
     \begin{subfigure}[b]{0.24\textwidth}
         \centering
         \includegraphics[width=0.8\textwidth]{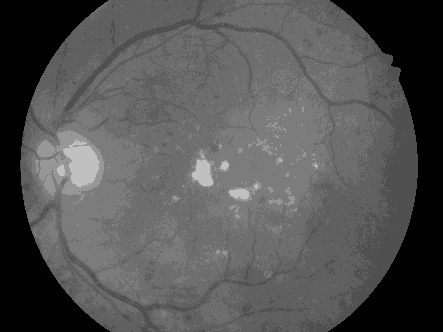}
         \caption{}
         \label{fig: fig4c}
     \end{subfigure}
     \hfill
     \begin{subfigure}[b]{0.178\textwidth}
         \centering
         \includegraphics[width=0.8\textwidth]{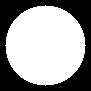}
         \caption{}
         \label{fig: fig4d}
     \end{subfigure}
     \begin{subfigure}[b]{0.2702\textwidth}
         \centering
         \includegraphics[width=0.8\textwidth]{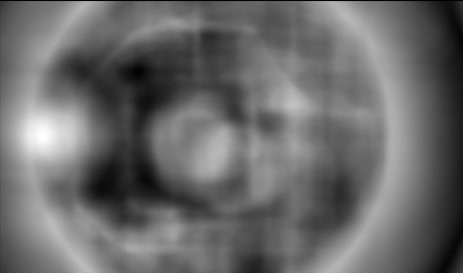}
         \caption{}
         \label{fig: fig4e}
     \end{subfigure}
     \begin{subfigure}[b]{0.24\textwidth}
         \centering
         \includegraphics[width=0.8\textwidth]{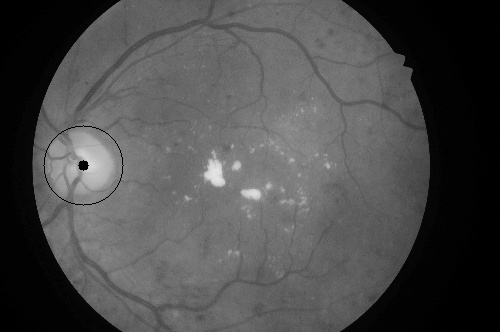}
         \caption{}
         \label{fig: fig4f}
     \end{subfigure}
     \begin{subfigure}[b]{0.24\textwidth}
         \centering
         \includegraphics[width=0.8\textwidth]{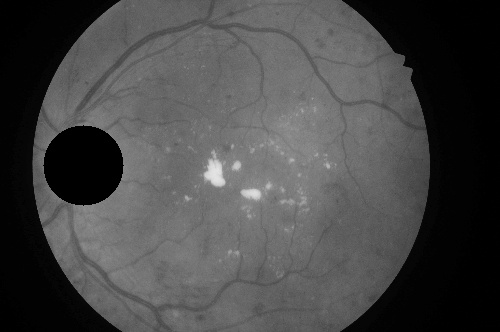}
         \caption{}
         \label{fig: fig4g}
     \end{subfigure}
        \caption{(a) Original image, (b) Grayscale of (a), (c) Result of k-means clustering, (d) Generated Template, (e) Template Matching result (using NCCOEFF; notice the OD region has highest similarity), (f) Marking OD and its center, (g) Masking OD region.}
        \label{fig: fig4}
\end{figure}

The proper localization of the OD is an essential step in the segmentation of exudates from fundus images, because of the comparable grey intensity levels of the OD and the exudates. The only difference is the irregular structure and (usually) smaller sizes of the exudates. The proposed method resized a coloured fundus image to 300 × 300 pixels and performed k-means clustering on its grayscale counterpart (Fig. \ref{fig: fig4c}). The k-means algorithm minimizes the squared difference between ‘k’ colour centres (or means) and the respective pixel values (colour values) in the image. If $x_i$ is the $i-th$ colour center and $p_j$ is the colour value of the $j-th$ pixel in the image, then the squared error function $J_{SE}$ to be minimized by k-means is given by:

\vspace{-1mm}
\begin{equation}
    J_{SE} = \sum_{i=1}^{k} \sum_{j=1}^{n} (||x_i - p_j||)^2
\end{equation}

A template of size comparable to the average size of the OD in all the images was generated (Fig. \ref{fig: fig4d}) and matched with the image, using the Normalized Correlation Coefficient (NCCOEFF) method (Fig. \ref{fig: fig4e}). NCCOEFF is computed as:

\vspace{-2mm}
\begin{equation}
 R(x, y) = \frac{{\sum}_{x', y'} ({T'^{(x',y')}} \cdot {I'^{(x+x',y+y')}})} {\sqrt{{\sum}_{x', y'} {T'{(x',y')^2}} \cdot {\sum}_{x', y'} {I'{(x+x',y+y')^2}}}}
\end{equation}

\begin{figure}
    \centering
    \includegraphics[width=1\textwidth]{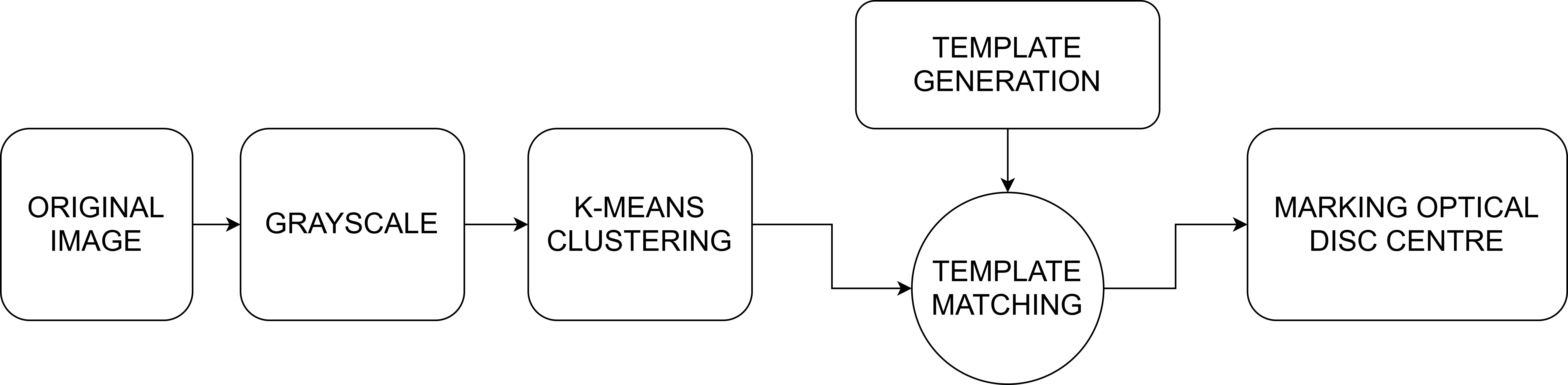}
    \caption{Optic Disc Localization Flow}
    \label{fig: fig5}
\end{figure}

The location of maximum values from the template-matched result was extracted. This indicates the approximate center of the Optic Disc in the image (Fig. \ref{fig: fig4f}). The algorithm's steps are shown in Fig. \ref{fig: fig5}.

\vspace{-2mm}

\vspace{-4mm}
\subsubsection{Detection of Exudates}

Exudates are small, yellowish deposits located on the outer layers of the retina, formed as a result of protein leakage from the retinal vessels.

\begin{figure}
     \centering
     \begin{subfigure}[b]{0.19\textwidth}
         \centering
         \includegraphics[width=\textwidth]{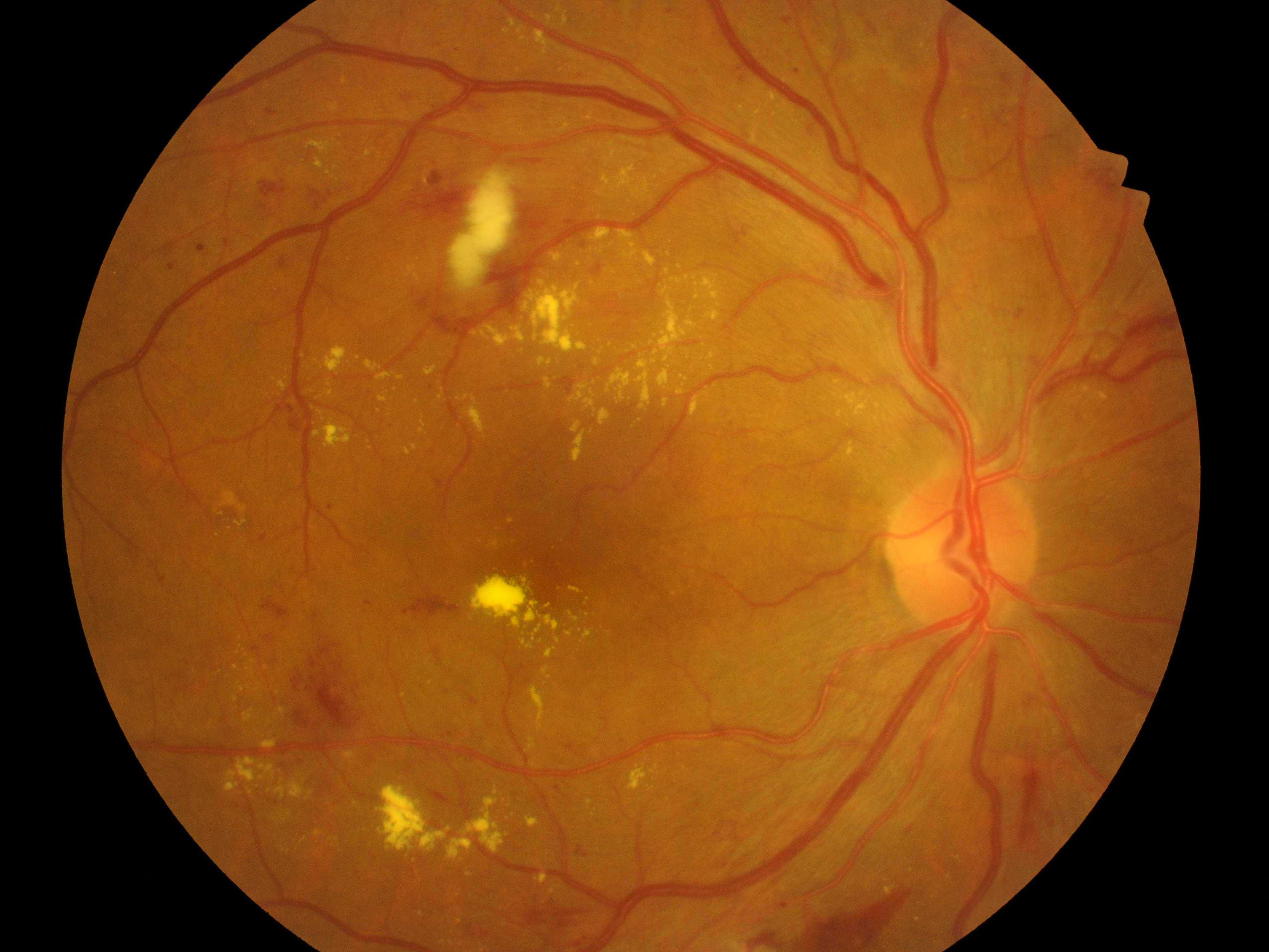}
         \caption{}
         \label{fig: fig6a}
     \end{subfigure}
     \hfill
     \begin{subfigure}[b]{0.19\textwidth}
         \centering
         \includegraphics[width=\textwidth]{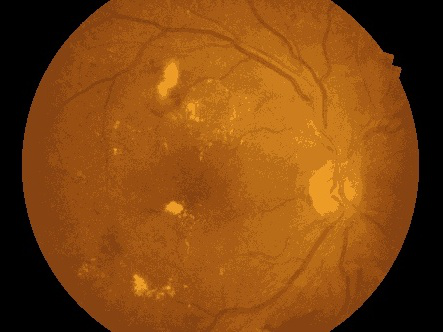}
         \caption{}
         \label{fig: fig6b}
     \end{subfigure}
     \hfill
     \begin{subfigure}[b]{0.19\textwidth}
         \centering
         \includegraphics[width=\textwidth]{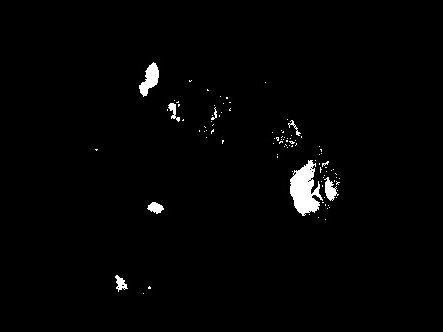}
         \caption{}
         \label{fig: fig6c}
     \end{subfigure}
     \hfill
     \begin{subfigure}[b]{0.19\textwidth}
         \centering
         \includegraphics[width=\textwidth]{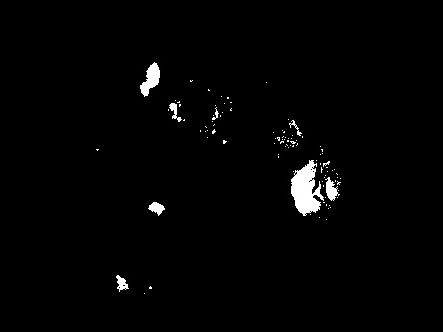}
         \caption{}
         \label{fig: fig6d}
     \end{subfigure}
     \begin{subfigure}[b]{0.19\textwidth}
         \centering
         \includegraphics[width=\textwidth]{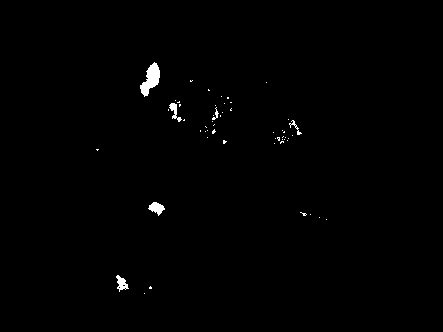}
         \caption{}
         \label{fig: fig6e}
     \end{subfigure}
     \caption{(a) Original Image, (b) K-means clustering result, (c) Extracting the exudates from (b) and thresholding, (d) Logical OR of (c) and the images containing the smallest exudates, (e) Final segmentation result after OD masking.}
     \label{fig: fig6}
\end{figure}

\begin{figure}
    \centering
    \includegraphics[width=1\textwidth]{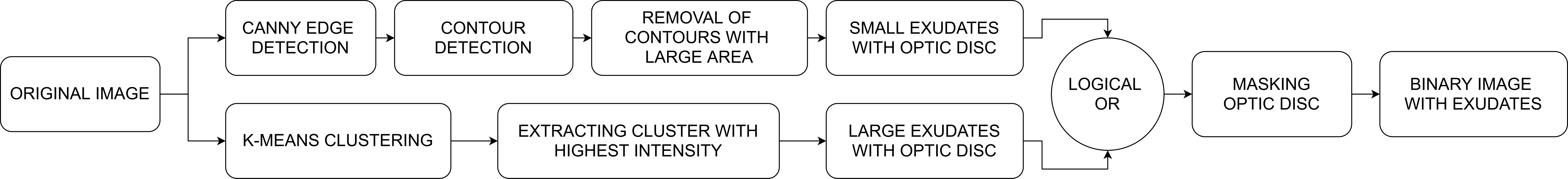}
    \caption{Exudates Detection Flow}
    \label{fig: fig7}
\end{figure}

Initially, k-means clustering was performed on the original image (Fig. \ref{fig: fig6b}), and the cluster with the highest intensities was extracted and binarized. This extracts the exudates (including the largest ones) along with the OD (Fig. \ref{fig: fig6c}). Canny Edge \cite{canny} and Contour detection were then sequentially performed on the green channel image to filter out large structures. Thresholding was applied to create a binary image, where the pixels with maximum grey level intensities were considered. This extracts the small exudates with the OD. The images containing the small and large exudates were then logically added (Fig. \ref{fig: fig6d}). The OD was detected in the grayscale image using the aforementioned method, and a circular, black mask was placed over it, in the image in Fig. \ref{fig: fig6d}. This yields the final segmentation result (Fig. \ref{fig: fig6e}). The algorithm's steps are shown in Fig. \ref{fig: fig7}.

\vspace{-4mm}

\subsubsection{Binary Diagnosis of Diabetic Retinopathy using Deep Convolutional Neural Network.}

The following DCNN was developed for the binary classification of DR.

\begin{figure}
    \centering
    \includegraphics[width=\textwidth]{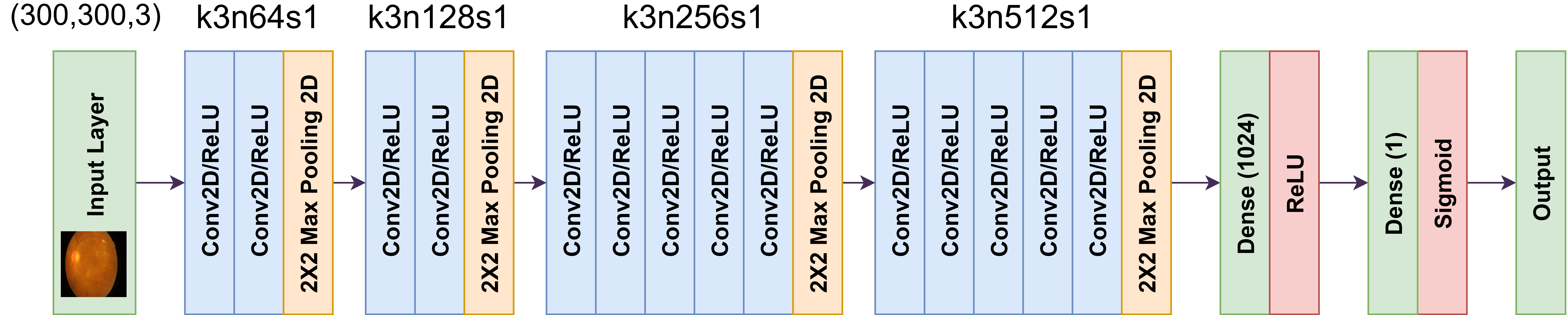}
    \caption{DCNN architecture with the corresponding kernel/filter sizes (k), number of feature maps (n) and strides (s) specified for each convolutional layer.}
    \label{fig: fig8}
\end{figure}

The proposed DCNN architecture was adapted from the VGG-16 \cite{vgg} architecture and improvised to perform DR diagnosis. It is composed of 4 stages of convolutional layers (Fig. \ref{fig: fig8}) with a 2x2 2D Max Pooling layer between each stage, which downsamples the input image by a factor of 2. In each convolutional layer, small 3×3 sized kernels were used with Rectified Linear Unit (ReLU) as the activation function.

\begin{equation}
    ReLU(x) = max(0,x),   \hspace{5mm}  x \hspace{1mm} \in \hspace{1mm} \mathbb{R}
\end{equation}

The original, coloured images are resized to 300×300 pixels, and trained with a batch size of 8. The output of the final convolutional stage is fed to a fully connected layer with 1024 neurons and the ReLU activation function. The final layer is a single neuron with the Sigmoid activation function for binary classification, which is given by:

\begin{equation}
    S(x) = \frac{1}{1+e^{-x}}
\end{equation}

To optimize the network, the Adam \cite{adam} optimizer was chosen having a learning rate of 0.0001 along with a Binary Cross-Entropy loss $J_{BCE}$, which can be computed as:

\begin{equation}
    J_{BCE} = - \frac{1}{m} \sum_{i=1}^{m} y_i \cdot log(p(y_i)) + (1-y_i) \cdot log(1-p(y_i))
\end{equation}

where $y_i$ is the actual label, p($y_i$) is the predicted probability of $y_i$ and $m$ is the number of training/test examples.

\section{Experimental Results}

\subsection{Segmentation of Blood Vessels}

\vspace{-2mm}
\begin{figure}[H]
    \centering
    \includegraphics[width=0.7\textwidth]{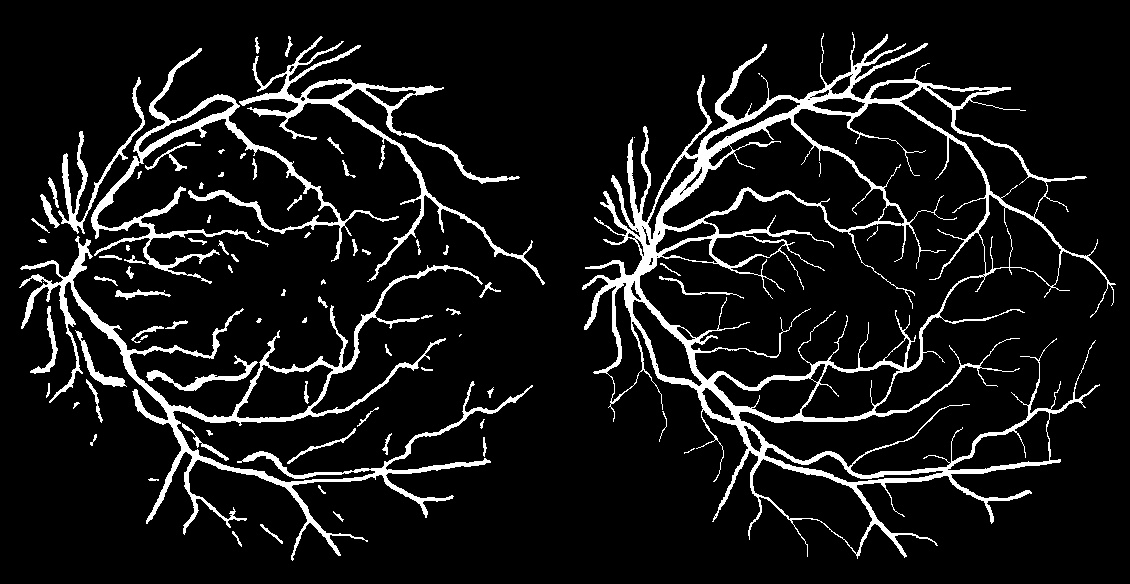}
    \caption{The segmentation result for the best case (left) with ground truth (right)}
    \label{fig: fig9}
\end{figure}

\vspace{-12mm}
\begin{table}[H]
    \centering
    \caption{Results of proposed vessel segmentation method on DRIVE dataset}
    \label{tab: table1}
    \vspace{2mm}
    \begin{tabular}{| p{0.24\textwidth} | p{0.24\textwidth} | p{0.24\textwidth} | p{0.24\textwidth} | }
    \hline
        \textbf{Accuracy(\%)} & \textbf{Specificity(\%)} & \textbf{Sensitivity(\%)} & \textbf{Dice Coefficient}  \\
        \hline
        95.93 & 98.32 & 71.19 & 0.75 \\
    \hline
    \end{tabular}
\end{table}

\subsection{Localization of Optic Disc}

The proposed Optic Disc Localization algorithm could accurately locate the OD in 80 of the 81 images in the whole dataset, achieving an accuracy of 98.77\%.

\begin{figure}
    \centering
    \includegraphics[width=0.7\textwidth]{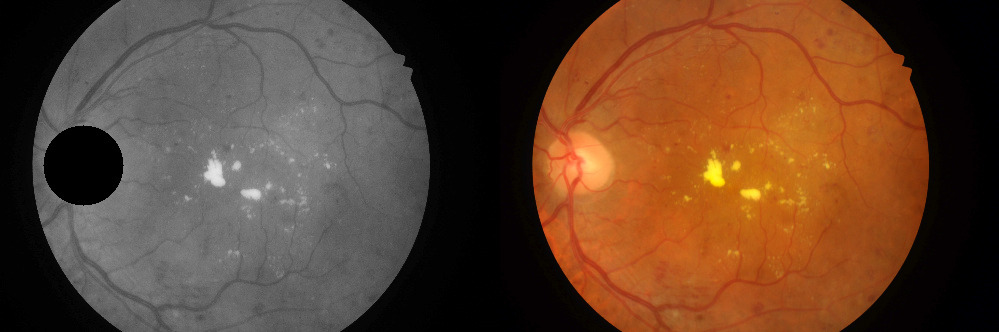}
    \caption{Optic Disc masked image (left) with the original image (right)}
    \label{fig: fig10}
\end{figure}

\subsection{Detection of Exudates}

The proposed algorithm was able to identify exudates in all the 81 images in the IDRiD Segmentation dataset quite appreciably.

\begin{figure}
     \centering
     \begin{subfigure}[b]{0.49\textwidth}
         \centering
         \includegraphics[width=\textwidth]{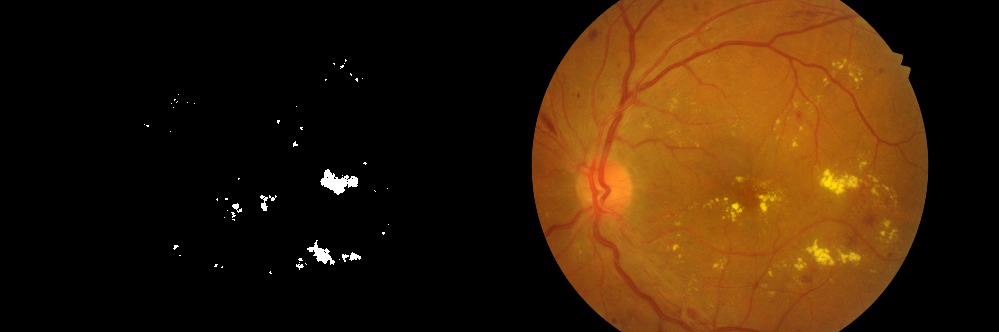}
         \caption{}
         \label{fig: fig11a}
     \end{subfigure}
     \hfill
     \begin{subfigure}[b]{0.49\textwidth}
         \centering
         \includegraphics[width=\textwidth]{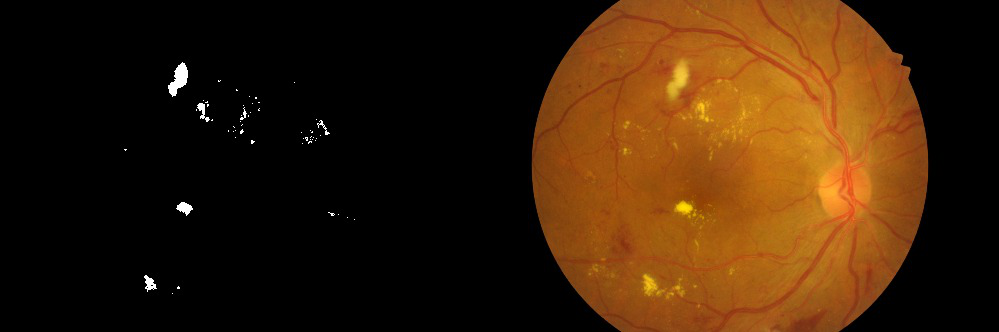}
         \caption{}
         \label{fig: fig11b}
     \end{subfigure}
     \caption{(a), (b) Result of proposed exudates detection method (left) with original image (right).}
     \label{fig: fig11}
\end{figure}

\subsection{Binary Diabetic Retinopathy Diagnosis}

The proposed model achieved a maximum training accuracy of 99.27\% over 100 epochs. The optimal accuracy on the test set was found to be 75.73\%.

\begin{figure}
    \centering
    \includegraphics[width=1\textwidth]{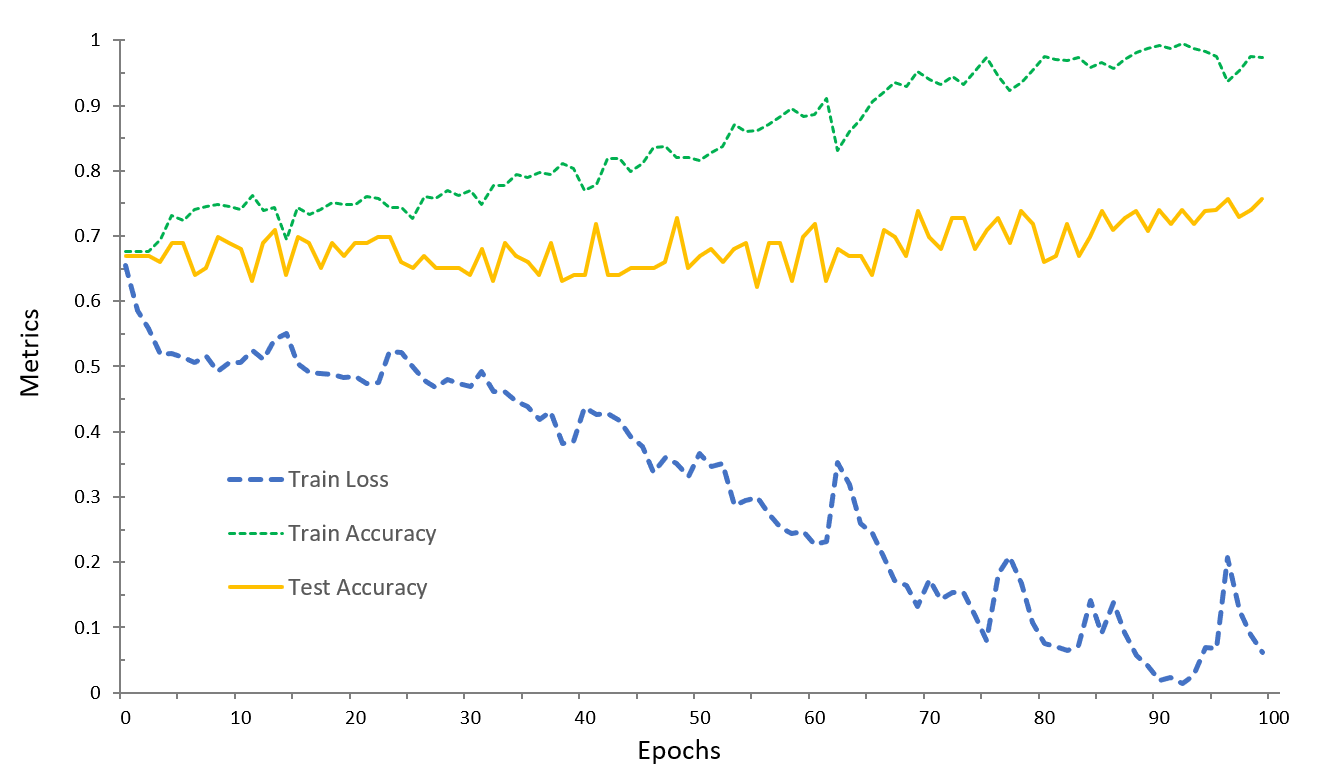}
    \caption{DCNN training results with Training Accuracy, Test Accuracy and Training Loss}
    \label{fig: fig12}
\end{figure}

\section{Conclusions}

The proposed blood vessel segmentation method performs better than most of the available segmentation techniques. It utilizes simple yet highly efficient morphological operations that perform background estimation and segmentation with more precision over existing methods. However, DCNNs are not as prone to noise and usually have higher sensitivity than image processing methods.

\begin{table}
    \centering
    \caption{Results of known vessel segmentation methods on DRIVE dataset (highest in bold).}
    \label{tab: table2}
    \vspace{2mm}
    \begin{tabular}{| p{0.30\textwidth} | p{0.07\textwidth} | p{0.2\textwidth} | p{0.19\textwidth} | p{0.19\textwidth} | }
    \hline
        \textbf{Method} & \textbf{Year} & \textbf{Accuracy (\%)} & \textbf{Specificity(\%)} & \textbf{Sensitivity(\%)}  \\
        \hline
        Azzopardi et al. \cite{azzo} & 2015 & 94.42 & 97.05 & 76.55 \\
        Roychowdhury et al. \cite{roychow} & 2015 & 95.20 & 98.30 & 72.50 \\
        GeethaRamani et al. \cite{geetha} & 2016 & 95.36 & 97.78 & 70.79 \\
        Christodoulidis et al. \cite{christo} & 2016 & 94.79 & 95.82 & \textbf{85.06} \\
        U-net \cite{r2unet} & 2018 & 95.31 & 98.20 & 75.37 \\
        R2U-net \cite{r2unet} & 2018 & 95.56 & 98.16 & 77.51 \\
        LadderNet \cite{laddernet} & 2018 & 95.61 & 98.10 & 78.56 \\
        BCDU-Net \cite{bcdunet} & 2019 & 95.60 & 97.86 & 80.07 \\
        Sun et al. \cite{sunx} & 2020 & 95.45 & 97.41 & 82.09 \\
        \emph{Proposed method} & 2020 & \textbf{95.93} & \textbf{98.32} & 71.19 \\
    \hline
    \end{tabular}
\end{table}

Methods that relied on the intensity levels of the OD, often fail to correctly locate it when the exudates have comparable grey level intensities. The proposed method is very efficient in locating the OD accurately, as it relies on the size and structure of the OD, instead of its grey level intensity. However, it may fail to distinguish between the OD and a large exudate, when the size of the latter is comparable to that of the OD.

Our exudates detection algorithm extracts not only the largest exudates but also the smaller ones because of the contour detection stage at the end. But the algorithm may generate false positives if the images are very unevenly illuminated.

The proposed DCNN architecture takes about an hour to train on Google Colab and has a test accuracy of 75.73\%, which is better than most of the previous architectures. However, the biggest advantage of the proposed architecture over existing networks is the significantly lower training time taken to produce acceptable results.

%
%

\end{document}